\documentclass[aps,prb,reprint,superscriptaddress]{revtex4-1}
\usepackage[dvips]{graphicx}

\begin{document}


\title{Spin-Orbital Kondo Effect in a Parallel Double Quantum Dot}


\author{Yuma Okazaki}
\affiliation{NTT Basic Research Laboratories, NTT Corporation, 3-1 Morinosato-Wakamiya, Atsugi, Kanagawa 243-0198, Japan}
\affiliation{Department of Physics, Tohoku University, Sendai, Miyagi 980-8578, Japan}
\author{Satoshi Sasaki}
\email[]{sasaki.s@lab.ntt.co.jp}
\affiliation{NTT Basic Research Laboratories, NTT Corporation, 3-1 Morinosato-Wakamiya, Atsugi, Kanagawa 243-0198, Japan}
\affiliation{Department of Physics, Tohoku University, Sendai, Miyagi 980-8578, Japan}
\author{Koji Muraki}
\affiliation{NTT Basic Research Laboratories, NTT Corporation, 3-1 Morinosato-Wakamiya, Atsugi, Kanagawa 243-0198, Japan}



\date{\today}

\begin{abstract}
Transport properties of the two-orbital Kondo effect involving both spin and orbital (pseudospin) degrees of freedom were examined in a parallel double quantum dot (DQD) with a sufficient interdot Coulomb interaction and negligibly small interdot tunneling. The Kondo effect was observed at the interdot Coulomb blockade region with degeneracies of both spin and orbital degrees of freedom. When the orbital degeneracy is lifted by applying a finite detuning, the Kondo resonance exhibits triple-peak structure, indicating that both spin and orbital contributions are involved.
\end{abstract}


\maketitle


The two-orbital Kondo effect is a many-body phenomenon that has been investigated for carbon-nanotube quantum dots (QDs),\cite{aJarillo-HerreroNature05,aChoiPRL05, aMakarovskiPRBR07, aAndersPRL08} vertical QDs,\cite{aSasakiPRL04} and parallel double quantum dots (DQDs),\cite{aWilhelmPhysica02, aHolleitnerPRB04, aHubelPRL08, aPohjolaEL01, aBordaPRL03, aGalpinJPhys06, aMravlijePRBR06, aOguchiJPSJ10, aChudonovskiyEPL05, aLimPRB06, aMartinsPhysica08, aKuboPRB08} where a localized state in the QDs has spin and orbital (pseudospin) internal degrees of freedom. Since the orbital degree of freedom plays the role of pseudospin in addition to spin, the Kondo correlation involves both spin and pseudospin flip events, and their strong mixing leads to spin-orbital Kondo properties that are in marked contrast to an ordinary SU(2) spin-Kondo effect, in which only spin flip events are involved.\cite{aGoldhaver-GordonNature98,aGoldhaver-GordonPRLO98, avanderWielScience00} The spin-orbital Kondo system is characterized by an interorbital interaction $V$ and intraorbital interaction $U$. Here $U$ ($V$) is associated with the spin (orbital) Kondo correlation, so the ratio $V/U$ is the critical parameter for developing a strong mixing of the spin and orbital. Experimentally, the spin-orbital Kondo effect has been observed in carbon-nanotube QDs\cite{aJarillo-HerreroNature05} and vertical QDs,\cite{aSasakiPRL04} where the spin and orbital satisfy the SU(4) symmetry because $U=V$ is satisfied. On the other hand, generally $V<U$ in parallel DQDs, which consists of two capacitively coupled QDs, and the two topmost levels in each QD form a two-orbital Kondo system. This asymmetry between $U$ and $V$ makes it difficult to observe spin-orbital Kondo effects in this system; indeed, experimental observation of such effect has not been reported.\cite{aWilhelmPhysica02, aHolleitnerPRB04, aHubelPRL08}

In this paper, we investigate the transport properties of the two-orbital Kondo effect in a parallel DQD in a region where both $V$ and $U$ are strong enough to enhance both spin and spin-orbital Kondo temperatures so that we can compare them. At a charge boundary between $(N+1,M)$ and $(N, M+1)$, where $N$ ($M$) denotes the electron number in the left (right) dot, the two states can be regarded as up and down pseudospin states that are energetically degenerate. We focus on the case with either or both of $N$ and $M$ being odd, where in addition to the orbital, the spin degree of freedom naturally comes into play. Along this charge boundary, where transport should be prohibited by interdot Coulomb interaction, we observe a Kondo-like enhancement of the conductance through the left dot characterized by a clear zero-bias peak (ZBP) and temperature scaling. As the orbital degeneracy is lifted by applying a finite detuning, the Kondo resonance splits into triple peaks, demonstrating that both spin and orbital contributions are involved. Our detuning-dependent data thus present a crossover from a spin-orbital Kondo state in the DQD to a spin-Kondo state in the left dot. 

\begin{figure}[b]
\includegraphics[scale=1]{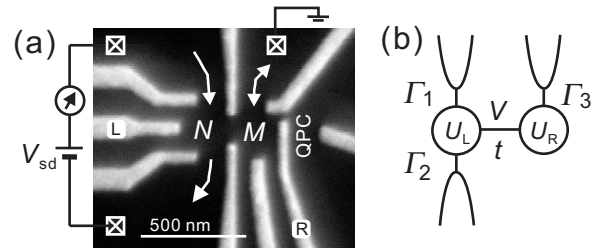}%
\caption{(a) A scanning electron microscope image of our device. (b) Schematic diagram of our parallel dot. $U_\mathrm{L}$ ($U_\mathrm{R}$) is on-site Coulomb interaction in the left (right) QD, $V$ and $t$ are interdot Coulomb interaction and tunneling, respectively. $\Gamma_1, \Gamma_2$ and $\Gamma_3$ are tunnel couplings to the leads. We estimate $U_\mathrm{L} = 650 \pm 50$\,$\mathrm{\mu eV}$, $t\sim 20 \pm 10$\,$\mathrm{\mu eV}$, $\Gamma = (\Gamma_1+\Gamma_2)\sim 260 \pm 30$\,$\mathrm{\mu eV}$, $\Gamma_3\sim \Gamma=260 \pm 30$\,$\mathrm{\mu eV}$, and $U_\mathrm{R}=1.2\pm 0.1$\,$\mathrm{meV}$ in the region of main focus in this study.}
\end{figure}

Figure 1(a) shows our device structure. A parallel DQD is defined in a two-dimensional electron gas with density $n=2.2\times 10^{11}$\,$\mathrm{cm^{-2}}$ and mobility $\mu=2\times 10^6$\,$\mathrm{cm^2/Vs}$ confined at a $\mathrm{GaAs/Al_{0.3}Ga_{0.7}As}$ interface 80\,nm below the surface. QDs are formed inside the two square areas enclosed by the gates, both of which are 220\,nm in lithographical size. The differential conductance $G=\mathrm{d}I/\mathrm{d}V$ through the left QD is measured using a standard lock-in technique with 3\,$\mu\mathrm{V}$ ac excitation at 78\,Hz. The right dot is tunnel-coupled to a single lead electrode kept at the ground and capacitively coupled to the left dot. All measurements were performed in a $\mathrm{^3He}$-$\mathrm{^4He}$ dilution refrigerator (the base temperature $T_\mathrm{b}\sim 40$\,mK). The number of electrons in the left (right) QD $N$ ($M$) is controled by gate voltage $V_{\mathrm{L}}$ ($V_{\mathrm{R}}$) applied to the gate L (R). The absolute values of $N$ and $M$ are determined by using a quantum point contact (QPC) charge detector. The device parameters are defined in Fig. 1(b).

\begin{figure}
\includegraphics[scale=1]{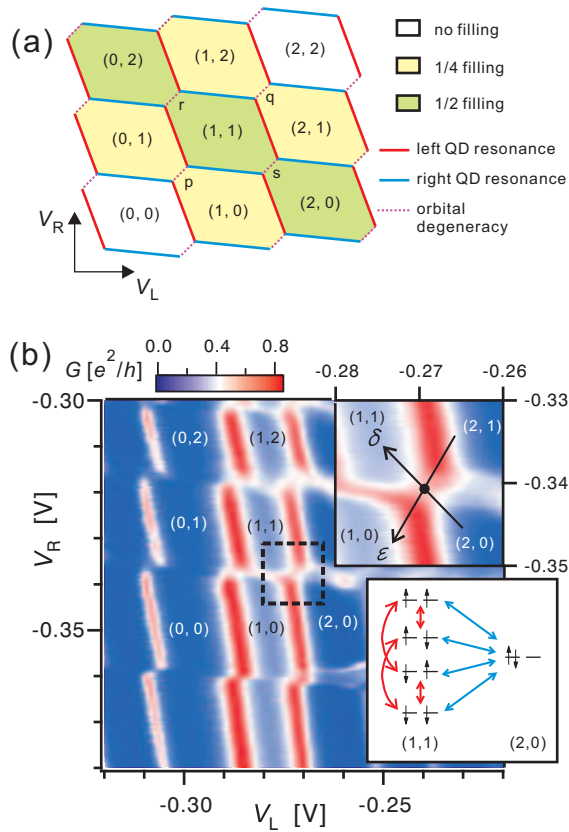}
\caption{(a) Charge stability diagram and the filling factors for the topmost levels as a function of gate voltages $V_\mathrm{L}$ and $V_\mathrm{R}$. The labels p, q, r, and s denote possible orbital degeneracy within these two orbitals. (b) Color plot of the measured differential conductance $G$ of the left QD. The number of electrons in the topmost levels are indicated in each region. Upper inset: Detailed measurement in the area enclosed by the rectangle in the main figure. Lower inset: Spin and orbital states of five-fold degeneracy at the boundary of $(1,1)/(2,0)$ occupancies.}
\end{figure}

The charge configuration of the DQD is determined by the on-site ($U_\mathrm{L}$, $U_\mathrm{R}$) and interdot ($V$) Coulomb interactions, which lead to a charge stability diagram exhibiting honeycomb structures as shown in Fig.~2(a). Due to on-site Coulomb blockade, a charge state $(N,M)$ is stable inside each hexagon. At charge boundaries between $(N, M)$ and $(N+1, M)$ [$(N,M)$ and $(N, M+1)$], the two adjacent charge states are energetically degenerate, which gives rise to a single-particle resonance peak in the transport through the left [right] QD. The lines defining the left and right QD resonances are disrupted by gaps that open as a result of interdot Coulomb interaction, where different classes of charge boundaries develop between $(N,M+1)$ and $(N+1,M)$, as denoted by p, q, r, and s in Fig.~2(a). Consequently, in these gap regions, tunneling through an individual QD is prohibited due to interdot Coulomb interaction, which we refer to as ``interdot Coulomb blockade''. At boundaries between $(N,M+1)$ and $(N+1,M)$, the two states are energetically degenerate, implying that cotunneling processes retaining the total electron number in the DQD are possible. Such cotunneling events lead to a pseudospin Kondo correlation, where the charge or orbital degree of freedom plays the role of a pseudospin. The orbital degeneracies associated with these two orbitals can be classified into two types depending on the filling factors: 1/4-filling [p and q in Fig.~2(a)] and 1/2-filling [r and s in Fig.~2(a)].\cite{aBordaPRL03,aOguchiJPSJ10} In the former, the spin and orbital degrees of freedom are independent, leading to four-fold degeneracy. In this case, an SU(4) Kondo effect is theoretically predicted to occur. In the latter, the four spin states associated with the (1,1) occupancy and one orbital state of (2,0) [(0,2)] occupancy form five-fold degeneracy, and the pseudospin-flip events for $(1,1)\leftrightarrow (2,0)$ [$(0,2)\leftrightarrow (1,1)$] can lead to a Kondo correlation [as shown in Fig.~2(b) lower inset for the case at (1,1)/(2,0) boundary]. 

Figure 2(b) shows the measured conductance through the left QD spanned by $V_\mathrm{L}$ and $V_\mathrm{R}$. The data comprise Coulomb peaks due to transport resonance through the left QD. The Coulomb peaks are disrupted each time the electron number in the right QD changes, indicating strong interdot Coulomb interaction in our DQD. Consequently, the charge stability diagram of the DQD is manifested by transport measurement through an individual QD. From the absolute number of electrons $(N, M)$ in the DQD determined using the QPC charge detector, we are able to identify the filling of the topmost orbital in each QD. With the scheme shown in Fig.~2(a), each hexagonal region is thus assigned its excess electron number $(n, m) = (N - 22, M - 6)$ as shown in Fig.~2(b). Here, (22, 6) is the number of inner-core electrons that do not contribute to the Kondo correlation in the relevant region. Consistent with the filling thus assigned, the on-site Coulomb blockade regions with $(n, m) = (1, m)$ show enhanced valley conductance due to the spin-Kondo effect in the left QD.

We here focus on the region near the interdot Coulomb blockade along the (1,1)/(2,0) boundary [indicated by the dashed rectangle in Fig.~2(b)], where one expects five-fold degeneracy associated with the spin and orbital degrees of freedom for this 1/2 filling. The upper inset of Fig.~2(b) depicts the result of a detailed measurement in this region. The data demonstrate Kondo-like conductance enhancement emerging along the (1,1)/(2,0) boundary, where transport should otherwise be prohibited by the interdot Coulomb blockade. As we detail below, we observe both ZBP and Kondo-temperature scaling for this feature, which corroborates the Kondo correlation underlying the observed conductance enhancement. The lower inset of Fig.~2(b) illustrates the possible spin configurations for each of the relevant charge states, (1,1) and (2,0). Arrows indicate possible spin and/or pseudospin flip processes that would contribute to the Kondo correlation.

\begin{figure}
\includegraphics[scale=1]{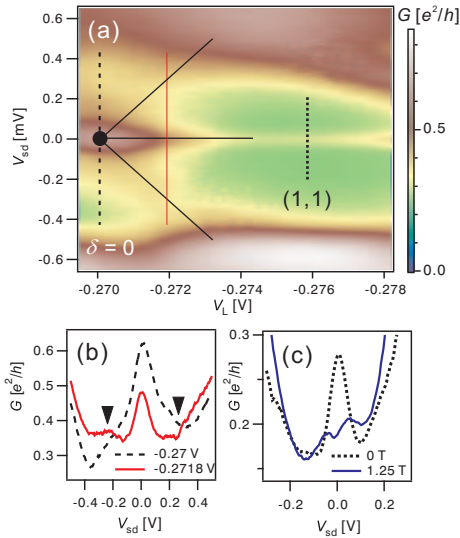}%
\caption{(a) Color plot of the differential conductance $G(V_{\mathrm{sd}})$, measured along the detuning axis $\delta$ defined in the upper inset of Fig.~2(b), plotted as a function of the corresponding gate voltage $V_\mathrm{L}$. (b) Detuning induced peak splitting of the Kondo resonance. Red solid (black dashed) line is extracted from (a) at $V_\mathrm{L}=-0.2718$\,$\mathrm{V}$ ($V_\mathrm{L}=-0.27$\,$\mathrm{V}$). (c) Zeeman splitting of the Kondo resonance measured at (1,1) valley center depicted by black dotted line in (a). Blue solid (black dotted) line is measured at $B=1.25$\,T ($B=0$\,T) .}
\end{figure}

\begin{figure}[t]
\includegraphics{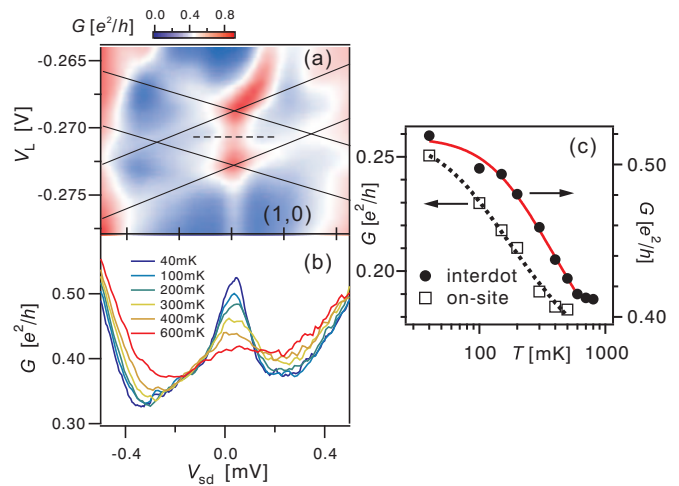}%
\caption{(a) Color plot of the source drain bias dependence of the differential conductance $G$ along $\epsilon$ axis defined in Fig.~2(b), as a function of the corresponding gate voltage $V_\mathrm{L}$. (b) Zero bias peak for different temperatures measured at the dashed line in (a). (c) Temperature dependence of the ZBP height at the interdot blockade (closed circles) and at (1,0) valley center (open squares) for comparison. The red solid and black dotted lines are the fitting results.}
\end{figure}

The contributions of pseudospin and spin to the observed Kondo effect can be identified by examining the dependence of the ZBP on the detuning and magnetic field. The detuning $\delta$, which is defined in the upper inset of Fig.~2(b), acts as an effective Zeeman energy for the pseudospin. Upon moving along the detuning axis from the (1,1)/(2,0) boundary toward the (1,1) valley center, one expects the (2,0) state to be lifted out from the five-fold degeneracy and the system to be left in four-fold degeneracy described as $\mathrm{SU(2)\otimes SU(2)}$. Figure 3(a) depicts the evolution of the measured conductance with the source-drain bias $V_\mathrm{sd}$ and the detuning $\delta$. The vertical dashed line in the figure marks the (1,1)/(2,0) degeneracy point (i.e., $\delta=0$). There are two important observations to be noted. First, a ZBP (i.e., a peak at $V_\mathrm{sd} = 0$ V) exists over the entire range of $\delta$ examined. Second, as indicated by the diagonal solid lines in the figure, there are two additional features splitting off from the ZBP at the (1,1)/(2,0) degeneracy point. The persistence of the ZBP to finite $\delta$ can be understood by recalling that, even at large $\delta$ where the system is purely in the (1,1) charge configuration, the spin degeneracy in each QD is retained and so the overall four-fold degeneracy of the DQD system is preserved. The spin contribution in the Kondo state in this regime can be identified by applying an in-plane magnetic field $B$. As shown in Fig.~3(c), applying $B = 1.25$ T at the (1,1) valley center [vertical dotted line in Fig.~3(a)] splits the ZBP into double peaks, with their separation corresponding to the Zeeman splitting $2|g|\mu_\mathrm{B}B$ ($g = -0.4$: g-factor of bulk GaAs. $\mu_\mathrm{B}$: the Bohr magneton).\cite{aKoganPRL04} On the other hand, the pseudospin contribution in the Kondo state at the (1,1)/(2,0) degeneracy point can be identified by applying a small detuning $\delta$. As shown in Fig.~3(b), we observe that the ZBP splits into triple peaks when a small $\delta$ is applied. (The split-off peak for positive $V_\mathrm{sd}$ is not well resolved because of the rising background conductance). Thus, the data in Fig.~3(a) demonstrate a crossover from a spin-orbital Kondo state involving the five-fold degeneracy of the DQD to an SU(2) spin-Kondo state in the left dot. (We note that a crossover from SU(4) to SU(2) Kondo states in a carbon-nanotube QD had been reported.\cite{aMakarovskiPRBR07}) It is interesting to note that, along with this crossover, the ZBP becomes narrower and its height smaller, suggesting the reduction in the Kondo temperature $T_\mathrm{K}$. 

We deduce the Kondo temperature by examining the temperature dependence of the ZBP. Figure 4(a) shows $V_\mathrm{sd}$ dependence of $G$ measured along the energy axis $\epsilon$ defined in the upper inset of Fig.~2(b). Here, we chose the $\epsilon$ axis (instead of the $\delta$ axis) in order to show the Coulomb diamond and the associated ZBP. As indicated by the solid lines in Fig. 4(a), a Coulomb diamond due to interdot Coulomb interaction, accompanied by a clear ZBP at its center, is visible. Figure 4(b) shows the ZBP for different temperatures, indicating that the ZBP is suppressed with increasing temperature. The closed circles in Fig.~4(c) depict the temperature dependence of the peak height. For comparison, the results for a similar measurement at the (1,0) on-site Coulomb blockade, where an ordinary SU(2) spin-Kondo effect occurs, are shown as open squares. We estimate the Kondo temperature using the following widely used empirical formula,\cite{aGoldhaver-GordonPRLO98}
$$G(T)=G_0 \left[1+(2^{1/s}-1)\left(\frac{T}{T_{\mathrm{K}}}\right)^{ 2}\right]^{-s}+G_1,$$
with $s=0.22$ the scaling parameter,\cite{com1} $G_0$ the zero temperature conductance $G(T=0)$, and $G_1$ the constant background. The solid and dotted lines in Fig.~4(c) represent the results of the fitting for the Kondo states in the interdot and on-site blockade regions, respectively. Here, we obtained $T_\mathrm{K}=0.9$~K with $G_1=0.27$~$e^2/h$ in the interdot blockade\cite{com2} and $T_\mathrm{K}=0.4$~K with $G_1=0.11$~$e^2/h$ in the on-site one, respectively.

There are two possible reasons for the larger $T_\mathrm{K}$ observed for the spin-orbital Kondo effect in the interdot blockade region: i) When spin and orbital degrees of freedom satisfy the SU(4) symmetry, their strong mixing enhances the Kondo effect and yields a higher $T_\mathrm{K}$, so-called the SU(4) Kondo effect.\cite{aJarillo-HerreroNature05} Similar Kondo enhancement may be possible even at the five-fold degeneracy where the SU(4) symmetry is no longer satisfied. ii) $T_\mathrm{K}$ depends on the energy $E$ of the single-particle states measured from the Fermi energy as $T_{\mathrm{K}}(V,\Gamma)\sim\sqrt{\Gamma V}\exp\left(-\pi E(E+U)/\Gamma U\right)/2$.\cite{aGoldhaver-GordonPRLO98, avanderWielScience00} This scaling formula gives the minimum $T_\mathrm{K}$ at the center of the blockade ($E=-U/2$), and $T_\mathrm{K}$ increases toward the Coulomb peak. At the interdot blockade, the single particle state is close to the Coulomb peak, and $T_\mathrm{K}$ is expected to be higher than that at the on-site blockade; From our measurements only, we cannot identify which mechanism is dominant.

So far we have focused on the interdot blockade region with 1/2 filling, where we have demonstrated a novel Kondo effect arising from the five-fold degeneracy associated with both spin and pseudospin. For the interdot blockade regions with 1/4 filling, where one expects an SU(4) Kondo effect, we did not observe a clear signature of a Kondo effect for the same set of dot tuning parameters. Combined with the fact that spin-Kondo features are clearly seen in the on-site Coulomb blockade regions with $(n, m) = (1, m)$, this suggests that in the present setup the tunnel coupling $\Gamma_3$ with the right lead is not sufficient for the pseudospin flip to be mediated by the right lead. We speculate that adjusting $\Gamma_3$ appropriately while keeping the strong interdot Coulomb interaction may allow us to observe an SU(4) Kondo effect. Further investigation is necessary to address this issue.

In conclusion, we investigated the two-orbital Kondo effect in the capacitively coupled parallel DQD with sufficient interdot coupling $V$ and negligible interdot tunneling $t$. We observed the Kondo effect at the interdot Coulomb blockade where spin and orbital degrees of freedom form five-fold degeneracy. The Kondo resonance was split into triple peaks at a finite detuning. From these results, we conclude that this Kondo effect involves both spin and orbital contributions.

We thank M. Eto, Y. Hirayama, T. Kubo, Y. V. Nazarov, R. Sakano, and Y. Tokura for valuable discussions.


\begin{thebibliography}{99}




\bibitem{aJarillo-HerreroNature05}P.~Jarillo-Herrero, J.~Kong, H.~S.~J.~van der Zant, C.~Dekker, L.~P.~Kouwenhoven, and S.~D.~Franceschi, Nature (London) \textbf{434}, 484 (2005).
\bibitem{aChoiPRL05}M.-S.~Choi, R.~L\'opez, and R.~Aguado, Phys. Rev. Lett. \textbf{95}, 067204 (2005).

\bibitem{aMakarovskiPRBR07}A.~Makarovski, A.~Zhukov, J.~Liu, and G.~Finkelstein, Phys. Rev. B \textbf{75}, 241407(R) (2007).

\bibitem{aAndersPRL08}F.~B.~Anders, D.~E.~Logan, M.~R.~Galpin, and G.~Finkelstein, Phys. Rev. Lett. \textbf{100}, 086809 (2008).

\bibitem{aSasakiPRL04}S.~Sasaki, S.~Amaha, N.~Asakawa, M.~Eto, and S.~Tarucha, Phys. Rev. Lett. \textbf{93}, 017205 (2004).

\bibitem{aWilhelmPhysica02}U.~Wilhelm, J.~Schmid, J.~Weis, and K.~v.~Klitzing, Physica E \textbf{14}, 385 (2002).

\bibitem{aHolleitnerPRB04}A.~W.~Holleitner, A.~Chudnovskiy, D.~Pfannkuche, K.~Eberl, and R.~H.~Blick, Phys. Rev. B \textbf{70}, 075204 (2004).

\bibitem{aHubelPRL08}A.~H\"ubel, K.~Held, J.~Weis, and K.~v.~Klitzing, Phys. Rev. Lett. \textbf{101}, 186804 (2008).


\bibitem{aPohjolaEL01}T.~Pohjola, H.~Schoeller, and G.~Sch\"on, Europhys. Lett. \textbf{54}, 241 (2001).

\bibitem{aBordaPRL03}L.~Borda, G.~Zar\'and, W.~Hofstetter, B.~I.~Halperin, and J.~v.~Delft, Phys. Rev. Lett. \textbf{90}, 026602 (2003).


\bibitem{aGalpinJPhys06}M.~R.~Galpin, D.~E.~Logan, and H.~R.~Krishnamurthy, Phys. Rev. Lett. \textbf{94}, 186406 (2005); J. Phys. Cond. Matt. \textbf{18}, 6545 (2006).

\bibitem{aChudonovskiyEPL05}A.~L.~Chudonovskiy, Euro. Phys. Lett. \textbf{71}, 672 (2005).

\bibitem{aMravlijePRBR06}J.~Mravlje, A. $\mathrm{Ram\check{s}ak}$, and T.~Rejec, Phys. Rev. B \textbf{73}, 241305(R) (2006).

\bibitem{aLimPRB06}J.~S.~Lim, M.-S.~Choi, M.~Y.~Choi, R.~L\'opez, and R.~Aguado, Phys. Rev. B \textbf{74}, 205119 (2006).


\bibitem{aKuboPRB08}T.~Kubo, Y.~Tokura, and S.~Tarucha, Phys. Rev. B \textbf{77}, 041305 (2008).

\bibitem{aMartinsPhysica08}G.~B.~Martins and C.~A.~B\"usser, Physica B \textbf{403}, 1514 (2008).

\bibitem{aOguchiJPSJ10}H.~Oguchi and N.~Taniguchi, J. Phys. Soc. Jpn. \textbf{79}, 054706 (2010).










\bibitem{aGoldhaver-GordonNature98}D.~Goldhaber-Gordon, H.~Shtrikman, D.~Mahalu, D.~Abusch-Magder, U.~Meirav, and M.~A.~Kastner, Nature (London) \textbf{391}, 156 (1998).

\bibitem{aGoldhaver-GordonPRLO98}D.~Goldhaber-Gordon, J.~Gores, M.~A.~Kastner, H.~Shtrikman, D.~Mahalu, and U.~Meirav, Phys. Rev. Lett. \textbf{81}, 5225 (1998).

\bibitem{avanderWielScience00}W.~G.~van der Wiel, S.~De Franceschi, T.~Fujisawa, J.~M.~Elzerman, S.~Tarucha, and L. P. Kouwenhoven, Science \textbf{5487}, 2105 (2000).

\bibitem{aKoganPRL04}A.~Kogan, S.~Amasha, D.~Goldhaber-Gordon, G.~Granger, M.~A.~Kastner, and H.~Shtrikman, Phys. Rev. Lett. \textbf{93}, 166602 (2004).


\bibitem{com1} This is the value known for the SU(2) Kondo effect [T.~A.~Costi and A.~C.~Hewson, J. Phys. Condens. Matter \textbf{6}, 2519 (1994)]. Due to the lack of a scaling theory for (1,1)/(2,0) degeneracy, we tentatively used the same $s$ value for the present case.

\bibitem{com2} Fitting with $s \sim 1$ and the same $G_1$ yields $T_\mathrm{K} \sim 0.7$~K. Both this value and $T_\mathrm{K}=0.9$~K for $s = 0.22$ are within a range consistent with the ratio of the ZBP width between the interdot blockade and the on-site one.

\end{thebibliography}
\end{document}